\documentclass[aps,prb,twocolumn,showpacs,amsmath,amssymb]{revtex4}

\usepackage{graphicx}
\usepackage{dcolumn}
\usepackage{bm}
\begin{document}

\title{Ultimate Informational Capacity of a Volume Photosensitive
\\ Media}

\author{Yuriy I. Kuzmin}
\email{yurk@mail.ioffe.ru}
\author{Viktor M. Petrov}
\email{vikpetroff@mail.ru}

\affiliation{Saint Petersburg State Polytechnical University, 29
Polytechnicheskaya Street, Saint Petersburg 195251 Russia}

\date{\today}

\begin{abstract}
The ultimate information capacity of a three-dimensional hologram
for the case of an optimal use of the dynamic range of a storage
medium, number of pages, the readout conditions is considered. The
volume hologram is regarded as an object of the information
theory. For the first time the formalism of the reciprocal lattice
has been introduced in order to estimate the informational
properties of the hologram. The diffraction-limited holographic
recording is analyzed in the framework of the reciprocal lattice
formalism. Calculations of the information capacity of a
three-dimensional hologram involve analysis of a set of
multiplexed holograms, each of which has a finite signal-to-noise
ratio determined by the dynamic range of the holographic medium
and the geometry of recording and readout. An optimal number of
pages that provides a maximum information capacity at angular
multiplexing is estimated.
\end{abstract}


\maketitle

\section{Introduction}

Information capacity and throughput of different optical objects,
including holograms, is of vital importance for optical
information processing and storage systems
\cite{heerden,ramberg,tiemann,petrov,wullert}. Here we analyze the
ultimate information capacity of a three-dimensional (volume)
hologram for the case of an optimal use of the dynamic range of
the storage medium.

\section{Information Capacity and Reciprocal Lattice Formalism}

Information capacity is directly related to the physical
properties of the optical media. It can be defined as the maximum
amount of information that can be recorded and then read out with
an arbitrarily small probability of error. According to the
Kotelnikov-Shannon sampling theorem, the information recorded in a
hologram is fully determined by $4A\Delta ^{2}W$ pixels, where $A$
is the hologram cross section, and $\Delta ^{2}W$ is the
two-dimensional width of the spectrum of recorded spatial
frequencies. The factor $4\Delta ^{2}W$ is a two-dimensional
analog of the Nyquist frequency. The upper limit of information
capacity of a three-dimensional and two-dimensional hologram can
be found by the analogy with the Shannon formula for the channel
capacity of a communication link in the presence of white noise
\cite{shannonbell,shannonire,fellgett,brady}.

\begin{equation}
C_{3D}=4A\Delta ^{2}WN\log _{2}\left( 1+R_{3D}\left( \Delta
^{2}W,N\right) \right) \text{\ , \ \ bit}  \label{cap3}
\end{equation}

\begin{equation}
C_{2D}=4A\Delta ^{2}W\log _{2}\left( 1+R_{2D}\left( \Delta
^{2}W\right) \right) \text{\ , \ \ bit}  \label{cap2}
\end{equation}%
where $N$ is the number of multiplexed holograms (pages),
$R=P_{s}/P_{n}$ is the signal-to-noise ratio at readout of one
pixel, $P_{s}$ is the upper boundary of the average power of the
image-forming signal, and $P_{n}$ is the average power of optical
noises.

\begin{figure}
\includegraphics{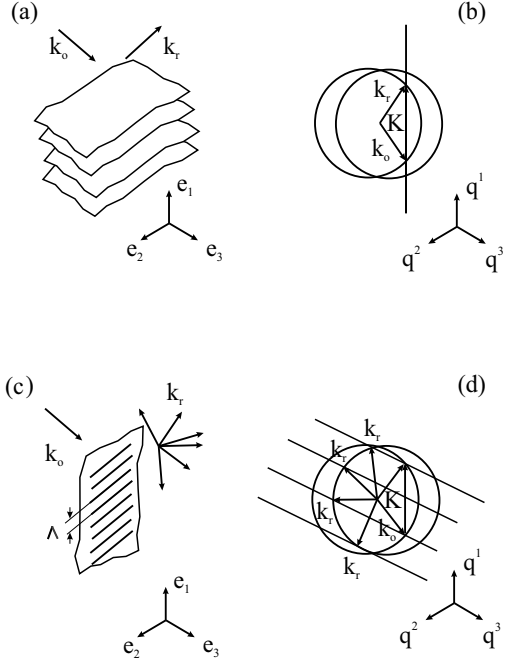}
\caption{\label{fig1} Translation-invariant sets and reciprocal
gratings for optical diffraction from three-dimensional (a, b) and
two-dimensional (c, d) holographic gratings: (a, c) - the
coordinate space, (b, d) - the \textit{k}-space; $\mathbf{K}$ is
the scattering vector (shown only for scattering \textquotedblleft
forward"), $\mathbf{k}_{0}$ and $\mathbf{k}_{r}$
are the wave vectors of the readout and reconstructed beams, respectively, $%
\Lambda =2\pi /\mathrm{K}$ is the grating period.}
\end{figure}

Let us find the maximum number of pixels that can be recorded in a
photosensitive media as a volume hologram in the case only
diffraction limitations exist. We assume that the elementary
holographic grating is a spatial distribution of the recorded
physical parameter invariant with respect to the translation of
the type.

\begin{equation}
\mathbf{T}_{3D}=n_{1}\mathbf{e}_{1}+\nu _{2}\mathbf{e}_{2}+\nu _{3}\mathbf{e}%
_{3}\text{\ \ , \ \ }\forall n_{1}\in \Im \ \text{,}\ \ \ \forall
\nu _{2},\nu _{3}\in \Re  \label{trans3}
\end{equation}

\begin{equation}
\mathbf{T}_{2D}=n_{1}\mathbf{e}_{1}+\nu _{2}\mathbf{e}_{2}+\nu _{0}\mathbf{e}%
_{3}\text{\ \ , \ \ }\forall n_{1}\in \Im \ \text{,}\ \ \ \forall
\nu _{2}\in \Re   \label{trans2}
\end{equation}%
for the three- and two-dimensional holograms, respectively; where $\mathbf{e}%
_{i}$ are the basis vectors of translation, $\nu _{0}$ is the
grating plane coordinate, $\Im $ and $\Re $ are the sets of
integer and real numbers. In the three-dimensional space the
vector of translation (\ref{trans3})
describes the set of parallel planes $\mathbf{T}_{3D}\in \Im \otimes \Re ^{2}$ (Fig.~%
\ref{fig1}, a), the vector in (\ref{trans2}) describes the set of
collinear and complanar lines $\mathbf{T}_{2D}\in \Im \otimes \Re $ (Fig.~\ref%
{fig1}, c).

The reciprocal lattices corresponding to translations (\ref{trans3}) and (%
\ref{trans2}) in the \textit{k}-space are

\begin{equation}
\mathbf{Q}_{3D}=m^{1}\mathbf{q}^{1}\text{\ \ , \ \ }\forall
m^{1}\in \Im \label{recip3}
\end{equation}

\begin{equation}
\mathbf{Q}_{2D}=m^{1}\mathbf{q}^{1}+\mu ^{0}\mathbf{q}^{2}+\mu ^{3}\mathbf{q}%
^{3}\text{\ \ , \ \ }\forall m^{1}\in \Im \ \text{,}\ \ \ \forall
\mu ^{3}\in \Re   \label{recip2}
\end{equation}%
where $\mu ^{0}$is the coordinate of the reciprocal lattice plane, $\mathbf{q%
}^{j}$ are the basis vectors of the reciprocal lattice satisfying
the orthogonality relation $\mathbf{e}_{i}\bullet
\mathbf{q}^{j}=2\pi \delta _{i}^{j}$, where $\delta _{i}^{j}$ is
the Kronecker symbol.

In the \textit{k}-space the reciprocal lattice (\ref{recip3}) is a
set of equidistant points $\mathbf{Q}_{3D}\in \Im $
(Fig.~\ref{fig1}, b) whereas the reciprocal lattice (\ref{recip2})
is a set of collinear and complanar lines $\mathbf{Q}_{2D}\in \Im
\otimes \Re $ , the orientation of which is defined by the
orthogonality relation (Fig.~\ref{fig1}, d).

When information is reading from a hologram, the orientation of
the reconstructed beam is determined by the points of intersection
of the Ewald
sphere (the radius of which is equal to the readout light wave vector $%
\mathbf{k}_{0}$) and the reciprocal lattice. The scattering vector
coincides
in this case with the vector of reciprocal lattice (\ref{recip3}) or (\ref%
{recip2}), as shown in Fig.~\ref{fig1}. Independent states in the \textit{%
k}-space at the Ewald sphere correspond to diffraction-resolvable
Fourier components of the reconstructed image. Therefore, the
maximum number of pixels that can be recorded in the hologram of
any dimension is equal to the number of states at 1/2 of the Ewald
sphere surface.

\begin{equation}
\sup \left( 4A\Delta ^{2}W\right) =\frac{1}{2}\left( \frac{4\pi k_{0}^{2}}{%
\Delta ^{2}k}\right) =8\pi \frac{A}{\lambda ^{2}}  \label{sup}
\end{equation}%
where $\Delta ^{2}k=\pi /A$ is the square of the minimum
uncertainty of the wave vector in the diffraction limit of
resolution, and $\lambda $ is the readout light wavelength.
Countingthe states on half the surface of the Ewald sphere
corresponds to summing over the spatial frequencies within the
entire Fourier plane. The expression (\ref{sup}) assumes that
there is no polarization multiplexing. If it is taken into
account, the results should be doubled.

\section{Dynamic Range, Word Length and Number of Multiplexed Pages}

The maximum number of pages recorded in a three-dimensional
hologram at angular multiplexing can be calculated by summation
over all vectors of the reciprocal lattice sitting on the Ewald
sphere (i.e., over all the wave vectors of the recorded
holographic gratings): $\max \left( N\right) =2k/\Delta
k=4L/\lambda $, where $\Delta k=\pi /L$ is the minimum uncertainty
of the wave vector, and $L$ is the hologram thickness. Taking into
account the expression (\ref{sup}), it is easy to find the maximum
number of pixels that potentially could be recorded at all pages
of a three-dimensional hologram in the case of an unlimited
dynamic range

\begin{equation}
\sup \left( 4A\Delta ^{2}W\right) \max \left( N\right) =32\frac{\pi AL}{%
\lambda ^{3}}  \label{supmax}
\end{equation}

Estimates of the type of $"volume"/\lambda ^{3}$ are often given
for the ultimate information capacity of a hologram
\cite{heerden,ramberg,wullert}; but an unjustified assumption of
information storage in the form of stored elements of volume
("voxels" \cite{fellgett}) is frequently made in this case, and
the two-dimensionality of the spectrum of spatial frequencies of
the recorded image is ignored. The number of multiplexed holograms
is determined by the finite dynamic range of the holographic
medium on which the signal-to-noise ratio depends. This is the
reason, while the estimation (\ref{supmax}) practically is
unachievable. At multiplexing, the information capacity does not
grow in $N$ times, as it could be
concluded from a shallow analysis of expressions (\ref{cap3}) and (\ref%
{supmax}) that did not take into account the dependence\
$R_{3D}=R_{3D}(N)$.

Let us consider now, how the number of pages affects the
information capacity. In the case of a two-dimensional hologram
the entire dynamic range is used to code each pixel with the
maximum word length. For a three-dimensional hologram an exchange
of the word length on the number of pages in the limits of the
same dynamic range is possible. An increase in the number of pages
is achieved by decreasing $R_{3D}$ up to the word length of one
bit per pixel. Let us show that there is an optimal number of
pages at which the information capacity is the highest. The number
of multiplexed
holograms can be presented in the form $N=\sqrt{P_{s}\left( 1\right) }/\sqrt{%
P_{s}}\left( N\right) $ , where $P_{s}(1)$ is the maximum signal
power for recording only one page for using the entire dynamic
range, $P_{s}(N)$ is the signal power for recording one of the $N$
multiplexed pages. Now we can relate $R_{3D}\left( N\right) $ and
$R_{3D}\left( 1\right) =\max_{N}R_{3D}\left( N\right) $ as

\begin{equation*}
R_{3D}\left( N\right) \equiv \frac{P_{s}\left( N\right) }{P_{n}}=\frac{%
P_{s}\left( 1\right) }{N^{2}P_{n}}=\frac{P_{3D}\left( 1\right)
}{N^{2}}
\end{equation*}

If $R_{3D}(1)>>N^{2}$ the expression (\ref{cap3}) acquires the
form

\begin{equation}
C_{3D}\left( N\right) =NC_{3D}\left( 1\right) -8A\Delta ^{2}WN\log
_{2}N \label{capfin}
\end{equation}

This function has a maximum at

\begin{equation*}
N_{0}=2\left( \frac{C_{3D}\left( 1\right) }{8A\Delta ^{2}W}-\frac{1}{\ln 2}%
\right)
\end{equation*}

Therefore, there is an optimal number of pages
$N_{opt}=entier(N_{0})$ above which the information capacity will
decrease because of a reduction in the signal-to-noise ratio
$R_{3D}(N)$.

It is interesting to note that in the case of a sufficiently high
$R_{3D}(1)$ the information capacity $C_{3D}(N)$ of the
three-dimensional hologram in
which $N$ pages are recorded is lower than the information capacity $%
NC_{3D}(1)$ of the set consisting of $N$ holograms in each of
which only one
page is recorded, all other conditions being equal. As follows from Eq. (\ref%
{capfin}), the difference in the information capacity per one
pixel is described by the function

\begin{equation*}
L\left( N\right) \equiv \frac{NC_{3D}\left( 1\right) -C_{3D}\left(
N\right) }{4A\Delta ^{2}W}=2N\log _{2}N\text{\  , \ \ bit}
\end{equation*}

\begin{figure}
\includegraphics{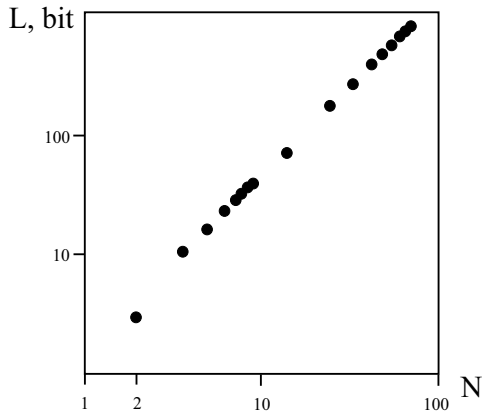}
\caption{\label{fig2} Function $L(N)$ versus number of multiplexed
holograms.}
\end{figure}

The numerical calculations are shown in Fig.~\ref{fig2}.

\section{Conclusion}

To summarize, the hologram was regarded in our study as an object
of the information theory. Calculations of the information
capacity of a three-dimensional hologram involved analysis of the
set of multiplexed holograms, each of which had a finite
signal-to-noise ratio determined by the dynamic range of the
storage medium. The problem of an optimal use of the dynamic range
at angular multiplexing was solved. Analysis of
diffraction-limited holographic information recording was carried
out in the framework of the reciprocal lattice formalism, which
allowed us to use such a basic property of an optical image as
two-dimensionality of the spectrum of its spatial frequencies to a
full extent.

\bibliography{hologram}

\begin{thebibliography}{9}
\expandafter\ifx\csname natexlab\endcsname\relax\def\natexlab#1{#1}\fi
\expandafter\ifx\csname bibnamefont\endcsname\relax
  \def\bibnamefont#1{#1}\fi
\expandafter\ifx\csname bibfnamefont\endcsname\relax
  \def\bibfnamefont#1{#1}\fi
\expandafter\ifx\csname citenamefont\endcsname\relax
  \def\citenamefont#1{#1}\fi
\expandafter\ifx\csname url\endcsname\relax
  \def\url#1{\texttt{#1}}\fi
\expandafter\ifx\csname urlprefix\endcsname\relax\def\urlprefix{URL }\fi
\providecommand{\bibinfo}[2]{#2}
\providecommand{\eprint}[2][]{\url{#2}}

\bibitem[{\citenamefont{Heerden}(1963)}]{heerden}
\bibinfo{author}{\bibfnamefont{P.~J.~V.} \bibnamefont{Heerden}},
  \bibinfo{journal}{Appl.\ Opt.} \textbf{\bibinfo{volume}{2}},
  \bibinfo{pages}{393} (\bibinfo{year}{1963}).

\bibitem[{\citenamefont{Ramberg}(1972)}]{ramberg}
\bibinfo{author}{\bibfnamefont{R.~G.} \bibnamefont{Ramberg}},
  \bibinfo{journal}{RCA\ Rev.} \textbf{\bibinfo{volume}{33}},
  \bibinfo{pages}{5} (\bibinfo{year}{1972}).

\bibitem[{\citenamefont{Tiemann et~al.}(2007)\citenamefont{Tiemann, Schmidt,
  Petrov, Petter, and Tschudi}}]{tiemann}
\bibinfo{author}{\bibfnamefont{M.}~\bibnamefont{Tiemann}},
  \bibinfo{author}{\bibfnamefont{M.}~\bibnamefont{Schmidt}},
  \bibinfo{author}{\bibfnamefont{V.~M.} \bibnamefont{Petrov}},
  \bibinfo{author}{\bibfnamefont{J.}~\bibnamefont{Petter}}, \bibnamefont{and}
  \bibinfo{author}{\bibfnamefont{T.}~\bibnamefont{Tschudi}},
  \bibinfo{journal}{Appl.\ Opt.} \textbf{\bibinfo{volume}{46}},
  \bibinfo{pages}{2683} (\bibinfo{year}{2007}).

\bibitem[{\citenamefont{Petrov et~al.}(1991)\citenamefont{Petrov, Stepanov, and
  Khomenko}}]{petrov}
\bibinfo{author}{\bibfnamefont{M.~P.} \bibnamefont{Petrov}},
  \bibinfo{author}{\bibfnamefont{S.~I.} \bibnamefont{Stepanov}},
  \bibnamefont{and} \bibinfo{author}{\bibfnamefont{A.~V.}
  \bibnamefont{Khomenko}}, \emph{\bibinfo{title}{Photorefractive crystals in
  coherent optical systems}} (\bibinfo{publisher}{Springer-Verlag},
  \bibinfo{address}{Berlin, New York}, \bibinfo{year}{1991}).

\bibitem[{\citenamefont{Wullert and Lu}(1994)}]{wullert}
\bibinfo{author}{\bibfnamefont{J.}~\bibnamefont{Wullert}} \bibnamefont{and}
  \bibinfo{author}{\bibfnamefont{Y.}~\bibnamefont{Lu}},
  \bibinfo{journal}{Appl.\ Opt.} \textbf{\bibinfo{volume}{33}},
  \bibinfo{pages}{2192} (\bibinfo{year}{1994}).

\bibitem[{\citenamefont{Shannon}(1948)}]{shannonbell}
\bibinfo{author}{\bibfnamefont{C.~E.} \bibnamefont{Shannon}},
  \bibinfo{journal}{Bell\ Syst.\ Tech.\ J.} \textbf{\bibinfo{volume}{27}},
  \bibinfo{pages}{379,623} (\bibinfo{year}{1948}).

\bibitem[{\citenamefont{Shannon}(1949)}]{shannonire}
\bibinfo{author}{\bibfnamefont{C.~E.} \bibnamefont{Shannon}},
  \bibinfo{journal}{Proc.\ IRE} \textbf{\bibinfo{volume}{37}},
  \bibinfo{pages}{10} (\bibinfo{year}{1949}).

\bibitem[{\citenamefont{Fellgett and Linfoot}(1955)}]{fellgett}
\bibinfo{author}{\bibfnamefont{P.~B.} \bibnamefont{Fellgett}} \bibnamefont{and}
  \bibinfo{author}{\bibfnamefont{E.~H.} \bibnamefont{Linfoot}},
  \bibinfo{journal}{Phil.\ Trans.\ Roy.\ Soc.} \textbf{\bibinfo{volume}{A247}},
  \bibinfo{pages}{369} (\bibinfo{year}{1955}).

\bibitem[{\citenamefont{Brady and Psaltis}(1992)}]{brady}
\bibinfo{author}{\bibfnamefont{D.}~\bibnamefont{Brady}} \bibnamefont{and}
  \bibinfo{author}{\bibfnamefont{D.}~\bibnamefont{Psaltis}},
  \bibinfo{journal}{J.\ Opt.\ Soc.\ Amer.\ A} \textbf{\bibinfo{volume}{9}},
  \bibinfo{pages}{1169} (\bibinfo{year}{1992}).

\end{thebibliography}

\end{document}